\documentclass[preprint,12pt]{elsarticle}

\usepackage{amssymb}
\usepackage{amsmath,amssymb,amsfonts}
\usepackage{algorithmic}
\usepackage{algorithm}
\usepackage{graphicx}
\usepackage{textcomp}
\usepackage{verbatim}
\usepackage{multirow}
\usepackage{tabu}
\usepackage{makecell}
\usepackage{color}
\usepackage{wrapfig}
\usepackage{url}
\usepackage{booktabs}
\usepackage{array}
\usepackage[hidelinks]{hyperref}

\usepackage{array}
\usepackage{multirow}
\usepackage{graphicx}

\journal{NPIC \& HMIT 2025}
\graphicspath{{figs_rds/}}

\begin{document}

\begin{frontmatter}

\title{Securing Radiation Detection Systems with an Efficient TinyML-Based IDS for Edge Devices}
\author[ot]{Einstein Rivas Pizarro\corref{cor1}}
\ead{einstein.rivaspizarro@ontariotechu.net}

\author[ot]{Wajiha Zaheer}
\ead{wajiha.zaheer@ontariotechu.net}

\author[ot]{Li Yang}
\ead{li.yang@ontariotechu.ca}

\author[ot]{Khalil El-Khatib}
\ead{Khalil.El-Khatib@ontariotechu.ca}

\author[ot]{Glenn Harvel}
\ead{Glenn.Harvel@ontariotechu.ca}

\address[ot]{Ontario Tech University, Oshawa, Ontario, Canada}

\begin{abstract}
\footnote{Preprint author original pre review. Accepted and Presented at NPIC \& HMIT 2025. The official proceedings version is available in the ANS Digital Library.}Radiation Detection Systems (RDSs) play a vital role in ensuring public safety across various settings, from nuclear facilities to medical environments. However, these systems are increasingly vulnerable to cyber-attacks such as data injection, man-in-the-middle (MITM) attacks, ICMP floods, botnet attacks, privilege escalation, and distributed denial-of-service (DDoS) attacks. Such threats could compromise the integrity and reliability of radiation measurements, posing significant public health and safety risks. This paper presents a new synthetic radiation dataset and an Intrusion Detection System (IDS) tailored for resource-constrained environments, bringing Machine Learning (ML) predictive capabilities closer to the sensing edge layer of critical infrastructure. Leveraging TinyML techniques, the proposed IDS employs an optimized XGBoost model enhanced with pruning, quantization, feature selection, and sampling. These TinyML techniques significantly reduce the size of the model and computational demands, enabling real-time intrusion detection on low-resource devices while maintaining a reasonable balance between efficiency and accuracy. 
\end{abstract}


\end{frontmatter}

\section{Introduction}

Nuclear materials play a paramount role in the well-being and advancement of modern society, contributing significantly to various sectors that enhance the quality of life, health, and technological progress. These materials are essential in energy production, medical applications, industrial processes, space exploration, and agriculture. They power homes and industries, advance medical treatments and diagnostics, ensure the safety and reliability of infrastructure, drive scientific research and innovation, enable space exploration, and support sustainable agriculture. However, the responsible and regulated operation of these materials, particularly in critical infrastructure services, requires significant effort, especially as the frequency and intensity of cyber-attacks continue to increase in such facilities \cite{r1,r2,r3,r4}. 

Recently, there has been a lot of anxiety about cyber threats that can target equipment related to safety and security functions and emergency-related support systems such as Critical Digital Assets (CDA) and Critical Systems (CS) \cite{r2,r3}. These critical components in radiation detection infrastructure, such as sensor networks and data transmission systems, present significant vulnerabilities for adversaries to exploit. Frequently, the initial point of compromise is through vulnerable IoT sensors in the detection network, which then facilitates the propagation of malicious code throughout the monitoring system and associated cloud infrastructure \cite{r1,r2,r4}. Other sophisticated attacks include data injection attacks, Distributed Denial-of-Service (DDoS), Man-in-the-Middle (MITM), flood attacks, botnet attacks, privilege escalation attacks, etc. \cite{r5,r6,r7}. These attacks can degrade the reliability of the system, leading to missed detections or false alarms, and impacting critical decision-making in radiation monitoring. Ensuring the integrity and security of RDSs is paramount for protecting public health and safety and for maintaining the effectiveness of radiation monitoring and response mechanisms. Additionally, Radiation Detection System (RDS) owners and custodians must incorporate sophisticated, multi-layered security solutions that leverage advanced technologies such as machine learning, artificial intelligence, and behavioral analytics to detect and respond to a wide range of evolving threats in real-time.

This paper focuses on the protection of RDSs, which are crucial components that help safeguard people and the environment by enabling the safe and secure use of civilian nuclear energy technologies and radioactive materials in facilities. The research, design, implementation, and testing of a Machine Learning (ML) -based Intrusion Detection System (IDS) is presented to enhance the security of these RDSs. Core to this endeavor is the application of TinyML techniques, which include feature reduction, undersampling, pruning, optimization, and quantization, to ensure the efficient and effective deployment of the IDS. TinyML enables real-time intrusion detection directly on edge devices with minimal computational overhead, reducing dependency on the cloud and mitigating potential attack surfaces in radiation detection networks \cite{r7}. The main contributions of this paper are as follows:
\begin{enumerate}
\item A novel RDS dataset was generated by applying K-Means clustering to group similar data points, calculating average feature values within each cluster for descriptive analysis, and introducing correlated noise and label noise to reduce overfitting and enhance realism.
\item A diverse range of common attacks is designed for behavioral profiling, classification, and anomaly detection tasks in the radiation detection sector.
\item A novel TinyML-based IDS using XGBoost and Optuna for parameter optimization is developed for resource-constrained environments. Enhanced by Open Neural Network Exchange (ONNX) for dynamic quantization, the proposed TinyML-based IDS transforms traditional ML models into efficient versions suitable for edge devices, maintaining high performance while reducing prediction time and model size.
\end{enumerate}

Currently, existing work lacks a systematic approach to reducing ML model complexity through feature selection, addressing data imbalance, and integrating TinyML with an optimization open-source framework, accompanied by a performance evaluation. This gap underscores the need for innovative solutions to enhance the security of RDSs.

The rest of this paper is organized as follows: Section 2 offers a comprehensive review of existing and related research in the field of radiation detection, with a particular emphasis on the application of TinyML in intrusion detection systems. Section 3 outlines the methodological framework, detailing a systematic approach to generating an ad-hoc RDS dataset, evaluating traditional ML models, employing AutoML techniques, and specifically focusing on TinyML methods to minimize model complexity. Section 4 presents a thorough performance evaluation, analyzing each technique and method utilized throughout the ML pipeline and discussing the final outcomes of the optimized ML model in real-world scenarios. Finally, Section 5 concludes the paper by summarizing the key findings of this research.
\section{Related Work}

\subsection{Overview of Existing Cybersecurity Advancements for RDSs}

In response to the evolving threats targeting the nuclear industry, significant efforts are being made to enhance the cybersecurity of nuclear facilities. A foundation of these efforts is the defense-in-depth strategy, which employs multiple layers of security controls to safeguard critical systems. This approach is highlighted in the NIST SP 800-82 standard \cite{r5}, focusing on the Cybersecurity of Industrial Control Systems (ICS). Defense-in-depth encompasses security management, physical security, network security, hardware security, and software security, ensuring that even if one layer is compromised, subsequent layers provide additional protection.

RDSs transmit signals and data to local or remote monitoring centers through various communication channels, enhancing threat detection and enabling timely responses to alarm conditions. However, data can be compromised during generation, processing, transmission, and display by external attackers, internal threats, or compromised devices \cite{r5}. Recent applications of advanced computational tools and techniques, such as reactor system design and analysis, plant operation and maintenance, and nuclear safety and risk analysis, have been implemented. However, existing models have limitations, including increased operational overhead and potential compatibility issues due to the complexity of integrating multiple security layers and technologies. 

Despite these challenges, advancements in Cybersecurity for RDSs provide a strong groundwork for protecting critical infrastructure against evolving threats. This research focuses on addressing these limitations and exploring innovative technologies, such as a TinyML model-based IDS potentially integrated into existing ML models to further enhance the security of RDSs. The proposed model extends its application beyond alarms generated by RDSs, potentially enhancing the security of sensors deployed for physical protection systems that secure other radioactive materials.

\subsection{TinyML Models for Cybersecurity}

TinyML represents a significant improvement in real-time monitoring and anomaly detection, particularly in the context of cybersecurity for RDSs \cite{r8}. By offloading complex computational tasks from sensor nodes to edge gateways, TinyML enables decision-making processes independent of cloud or remote computing infrastructure. This decentralization enhances the responsiveness and reliability of anomaly detection systems, making them less vulnerable to disruptions in connectivity.

Recent research has shown promising results in leveraging TinyML to counteract threats and attacks. Similarly, Li \textit{et}\textit{ al.} \cite{r9} explored various techniques within the data pre-processing and feature selection stages, demonstrating improved performance through metrics such as increased accuracy, reduced training time, and higher Area Under the Curve (AUC). Feature engineering and model optimization-based classification methods for network intrusion detection \cite{r10} discussed the use of random undersampling, lightweight model frameworks, and real-time environment testing. Yang \textit{et al.} \cite{r11} provided definitions and insights into AutoML techniques, while other studies \cite{r12} compared TinyML suites, focusing on pruning and quantization techniques. 

While previous studies have demonstrated the potential of ML models for intrusion detection \cite{r9} \cite{r10}, research on applying TinyML techniques for this purpose remains relatively unexplored. Existing approaches often focus on data pre-processing and feature engineering, improving ML performance metrics but neglecting optimization techniques such as pruning and quantization to reduce model complexity. In contrast, this research introduces several novelties to address the above limitations. We demonstrate the promising potential of integrating techniques such as pruning, quantization methods, and other pre-processing and feature engineering techniques with AutoML. Additionally, we trained and tested our ML models using an industry-specific dataset generated specifically for this domain. These integrations serve as enablers for efficient and practical deployment on edge devices. 

\section{Methodology}
To achieve our objective of reducing the complexity of a ML model and enabling TinyML to detect potential network attacks on resource-constrained devices at the edge of critical infrastructure, we have designed and developed a sequence of methods and techniques to optimize a supervised ML model using our novel, tailor-made RDS dataset. Our approach extends traditional ML models by innovatively reducing time complexity and memory consumption. This is accomplished through several key steps: addressing class imbalance, performing feature selection, conducting hyperparameter tuning, and subsequently incorporating pruning and quantization methods. By leveraging efficient algorithms and model optimization techniques, we enhance the suitability of our model for edge and embedded devices, with only marginal loss in classifier performance. These techniques are essential for deploying ML models on resource-constrained devices, ensuring that the models operate effectively within limited memory, computational power, and energy constraints.
\subsection{Data Generation}

We created a tailored RDS dataset using data from the bGeigieZen \cite{r13}, a device designed to accurately measure and map radiation levels. The methodology employed to produce this new dataset significantly builds upon the “synth-rad-data-gen-utils” project \cite{r14}, which is publicly available on GitHub. This project provides a suite of utilities for generating synthetic background radiation time series data, simulating real radiation sensors, and simulating cyber-attacks on radiation detection scenarios. By incorporating a broader variety of attack families, this effort has yielded a novel source of more realistic and diverse datasets, thereby enhancing the training of our ML model to better protect RDSs.

We began our work by importing an original dataset from Safecast, which included features such as timestamps for captured and uploaded times, geographic coordinates, radiation levels, and various metadata. The dataset initially underwent a series of pre-processing steps to ensure data integrity and prepare it for analysis. Following this, we identified and dropped irrelevant or non-numeric features, such as ‘Unit’, ‘Location Name’, ‘MD5Sum’, ‘Height’, ‘Surface’, ‘Radiation’, ‘Loader ID’, and ‘Device ID’, as they were deemed unnecessary for the clustering analysis. We retained numerical features, including ‘Captured Time’, ‘Latitude’, ‘Longitude’, ‘Value’, and ‘Uploaded Time’, for further processing. A pre-processing pipeline was created to handle missing values and normalize the numerical features using a SimpleImputer and MinMaxScaler, respectively. This pipeline was applied to the dataset using a column transformer, which also dropped the irrelevant features.

The pre-processed dataset was then subjected to clustering using the K-Means algorithm, an unsupervised learning method, to group similar data points into clusters. The number of clusters was determined based on the data characteristics and through experimentation. The K-Means model was fitted to the dataset, and each data point was assigned to a cluster. The clustering results were saved and analyzed to understand the distribution of samples across clusters. Descriptive statistics, such as the average value of each feature within each cluster, were calculated to provide insights into the characteristics of each cluster.

To simulate various types of cyber-attacks on RDSs, we defined attack generation functions based on the clustering results. These functions altered the dataset to mimic different attack scenarios, including data injection, man-in-the-middle (MITM) attacks, ICMP floods, botnet attacks, privilege escalation, and distributed denial-of-service (DDoS) attacks \cite{r5,r6,r7}. Data injection attacks could potentially manipulate sensor readings to report inaccurate radiation levels or falsified critical log events. Similarly, MITM attacks can subtly alter transmitted data, making modifications difficult to detect while misleading monitoring systems and decision-making processes. In addition, ICMP floods might be used by attackers to introduce delays to the uploaded time, disrupting real-time monitoring without altering measurements. Botnet attacks could simulate coordinated anomalous data from compromised components, creating inconsistencies in the dataset. Privilege escalation attacks directly modify system or administrative parameters, simulating unauthorized access by attackers with elevated permissions. DDoS attacks can overwhelm the system with excessive data, causing significant reporting delays and disrupting real-time analysis. These attacks collectively degrade the system’s reliability, leading to missed detections or false alarms and impacting critical decision-making in radiation monitoring \cite{r15}.

For each function, specific modifications were introduced to the ‘Value’ and ‘Uploaded Time’ features to reflect the impact of the respective attack. The attack functions were applied to specific clusters identified through the K-Means results. Finally, we encoded the features using the label encoding method to convert categorical variables into numeric format, ensuring compatibility with ML algorithms. Missing values were filled with zeros to maintain data integrity. To simulate realistic conditions, we introduced label noise by randomly flipping a fraction of the labels, mimicking labeling errors in real-world data.

This systematic approach ensured the creation of a comprehensive and tailored RDS dataset, prepared for further analysis and the development of robust cybersecurity measures. Initial attacks on RDSs involve strategies aimed at compromising data integrity and reliability, such as manipulating sensor readings, disrupting data transmission, or overwhelming the system. Understanding these attack methods is crucial for developing effective defense mechanisms and ensuring the safety and accuracy of RDS. This dataset serves as a critical foundation for developing and evaluating robust cybersecurity measures for RDSs, due to its diverse class distribution, balanced representation of attacks, and wide range of cyber threats.

\subsection{ML Model Development}

This work focuses on traditional ML algorithms that employ ensemble learning methods by combining multiple weak learners to create a strong learner. Specifically, Random Forest (RF), XGBoost, LightGBM, and CatBoost are evaluated for initial model selection to identify the most effective model for our tailored RDS dataset. These supervised learning models offer a balance of efficiency, performance, and robustness, making them well-suited for the unique challenges posed by our dataset.

RF \cite{r7} is a tree-structured ML model that uses a collection of decision trees for classification and regression. Known for its resilience and reliability on non-linear and complex datasets, it is widely applied in IDSs. XGBoost \cite{r16} is designed to improve speed and performance in ML tasks involving structured/tabular data. It iteratively combines decision trees to minimize prediction error using an additive model framework. Incorporates enhancements such as regularization, parallelization, and sparsity-aware algorithms for handling missing values efficiently. LightGBM \cite{r16} is recognized for its faster training speed, lower memory usage, and superior performance on large datasets. It uses histogram-based algorithms to bucket continuous features, leading to reduced memory usage and faster split calculations. Its leaf-wise growth strategy allows for faster convergence and often results in more accurate models. CatBoost \cite{r16} is designed to handle datasets with a significant proportion of categorical features. It introduces a novel approach for transforming categorical features directly during training using a combination of one-hot encoding and order-based encodings. It employs symmetric tree structures to simplify model complexity, improving inference speed and enabling compatibility with low-latency environments.

These algorithms are valuable for real-time multi-class classification tasks on resource-constrained devices and are widely used in IoT data analytics \cite{r10}. After the initial evaluation of these ML models, their TinyML versions are assessed and compared to determine the most effective and efficient final prediction model. The main goal is to identify the model that best balances resource usage and performance among these algorithms.

\subsection{TinyML Techniques}

The application of TinyML is a critical component of this research. By using TinyML methods \cite{r7}, this rapidly evolving field focuses on enabling the deployment and execution of ML models to detect cybersecurity attacks through inference \cite{r8,r10,r11}. However, data pre-processing and feature engineering are essential steps when implementing ML models directly on edge devices. Techniques such as data balancing through under-sampling, feature selection, pruning, and quantization effectively enhance the capability of ML models to operate on resource-constrained industrial components.

\subsection{Feature Selection}

Feature selection is a critical pre-processing step in ML that identifies the most relevant features and removes redundant ones to improve model performance. By reducing the dimensionality of the dataset while preserving essential information, feature selection mitigates computational inefficiencies and addresses the “curse of dimensionality,” where the number of samples required for accurate modeling grows exponentially with the number of features \cite{r10}. In addition to reducing computational costs, feature selection enhances model performance by eliminating noisy, irrelevant, and redundant features that can obscure decision-making and lead to overfitting \cite{r17}. Consequently, feature selection not only simplifies data complexity but also ensures robust and efficient model training.

In this research, we applied the SelectKBest method with the ANOVA F-value to perform feature selection \cite{r9}. This technique selects the top k features based on their statistical significance, measuring the linear dependency between each feature and the network activity classification, which is the target variable. Our implementation involved iteratively evaluating the performance of RF using different feature subsets selected by SelectKBest. Early stopping criteria based on relative improvement ensured a balanced feature subset. The final selected features formed a reduced dataset for further analysis and modeling. Feature selection contributes to our objective of reducing model complexity by decreasing the memory footprint required for training and prediction, as well as the time taken by the model to perform these tasks. 

\subsection{Sampling}

Undersampling is a widely used technique to address class imbalance by reducing the size of majority classes while retaining all minority class instances \cite{r18}. This method balances the dataset, improving the model’s ability to generalize from minority classes, which are often critical in applications such as anomaly detection and intrusion detection. In domains like radiation monitoring, where normal operations constitute the majority of the data and attacks represent a minority, undersampling mitigates the risk of missed detections and false alarms that could undermine system reliability. Furthermore, by reducing dataset size, undersampling lowers computational requirements, making it particularly advantageous for deployment in resource-constrained environments.

To implement undersampling, the majority class samples were reduced to match the average count of minority classes using a resample function. This process ensured a balanced dataset while preserving its key characteristics. By effectively addressing class imbalance, this methodology enables ML models to remain robust and efficient in handling multi-class classification tasks in imbalanced datasets. Moreover, since the primary objective of the proposed framework is to balance model effectiveness and efficiency,under-sampling methods are more suitable than oversampling techniques, as they reduce dataset size and model complexity.

\subsection{Pruning}

Pruning is a critical technique in ML used to reduce model complexity by removing parts that contribute little to predictive power \cite{r19}. This process not only simplifies the model but also enhances its interpretability and efficiency by preventing overfitting and reducing resource consumption. In contexts where real-time processing and resource constraints are critical, such as embedded systems, pruning ensures that models remain lightweight while maintaining strong predictive performance. By focusing on the most informative splits and discarding redundant or insignificant branches, pruning enables ML  models to strike a balance between accuracy and computational efficiency.

To perform pruning, hyperparameters that govern the selected tree-based ML models’ complexity,such as the number of trees, leaves, maximum depth, minimum samples per leaf, and subsample ratios,are systematically optimized. The process begins with defining an objective function that evaluates both performance metrics (e.g., F1-score, accuracy) and resource utilization (e.g., inference time, memory usage). During optimization, these hyperparameters are adjusted iteratively to identify configurations that minimize computational costs while maintaining accuracy above a predefined threshold. Multi-objective optimization is employed to explore trade-offs between competing goals, such as maximizing predictive performance and minimizing execution time. The results from this optimization process are analyzed to derive a simplified yet robust model configuration that is well-suited for deployment in resource-constrained environments.

\subsection{Quantization}

Quantization is a technique used to optimize ML models by reducing the numerical precision of their parameters, such as converting floating-point values to lower-bit integers. This process minimizes the model’s memory footprint and computational demands while preserving its predictive performance, making it essential for deploying models in resource-constrained environments like embedded systems. By reducing memory requirements and accelerating inference times, quantization enables efficient deployment on edge devices without compromising reliability, even for latency-sensitive applications such as real-time anomaly detection \cite{r19,r20}.

To implement quantization, we converted the trained model (.pkl) into a standardized format (ONNX) for interoperability across platforms. The model’s parameters were then dynamically quantized, reducing their precision to optimize memory usage and computational efficiency. This step involves calibrating the quantization process to minimize accuracy loss, validated through rigorous evaluation of performance metrics (e.g., F1-score, precision) and computational efficiency (e.g., inference time, cell execution time, and memory usage). The quantized model was tested to verify its suitability for deployment in hardware-constrained environments, ensuring a balance between efficiency and predictive power. 

\section{Performance Evaluation}

This section presents a review of the experimental results, highlighting the significant improvements in computational and time complexity reduction achieved through the application of TinyML methods. The results clearly demonstrate the potential of deploying ML models on very limited edge devices or embedded components operating in the closest layer to sensing and detection, effectively addressing intrusion detection threats.

\subsection{Experimental Setup}

The entire modeling process was conducted on a Microsoft Windows 11 Pro system (version 10.0.22631 Build 22631) equipped with an Intel(R) Core(TM) i7-8665U CPU @ 1.90GHz, 2112 MHz, featuring 4 cores and 8 logical processors, and 16 GB of RAM. The technology stack used for designing, fitting, training, and testing the models included open-source ML tools such as Jupyter Notebook (version 4.2.5) as the development environment (IDE), Python (version 3.12.4) as the programming language, Anaconda (MSC v.1929 64 bit (AMD64)) as the package manager, Conda (version 24.5.0) for the virtual environment, and Git (version 2.32.0) for version control. Additionally, the following libraries and frameworks were utilized: NumPy (version 1.21.2), Pandas (version 1.3.3), Scikit-learn (version 1.5.1), XGBoost (version 2.1.1), LightGBM (version 4.5.0), Matplotlib (version 3.4.3), Optuna (version 2.9.1), Tensor Flow (version 2.18.0), PyTorch (version 2.5.1), and PSUtil (version 5.8.0).

\subsection{Results and Analysis}

The initial evaluation encompassed five ML models: RF, XGBoost, LightGBM, CatBoost, and Long Short-Term Memory (LSTM). The LSTM model is evaluated as a baseline model for performance comparison. Table 1 presents the detailed performance metrics.

\begin{table}[t]
\centering
\caption{Initial comparison of five ML models}
\label{tab:init-comparison}
\renewcommand{\arraystretch}{1.2}
\scalebox{0.77}{
\begin{tabular}{|>{\centering\arraybackslash}m{5.5em}|%
                >{\centering\arraybackslash}m{4.5em}|%
                >{\centering\arraybackslash}m{4.5em}|%
                >{\centering\arraybackslash}m{4.5em}|%
                >{\centering\arraybackslash}m{6em}|%
                >{\centering\arraybackslash}m{5em}|%
                >{\centering\arraybackslash}m{5em}|}
\hline
\multirow{2}{*}{\textbf{Model}} &
\multicolumn{2}{c|}{\textbf{Performance}} &
\multicolumn{4}{c|}{\textbf{Computational Performance}}\\ \cline{2-7}
& \textbf{Accuracy (\%)} & \textbf{F1 score (\%)} & \textbf{Inference time (sec)} & \textbf{Inference time (ms/sample)} & \textbf{Memory usage (KB)} & \textbf{Cell execution time (sec)}\\ \hline
RF        & 99.895 & 99.895 & 0.3181  & 0.0104 & 135.3506  & 0.3181  \\ \hline
XGBoost   & 99.835 & 99.835 & 0.2684  & 8.8357 & 274.6221  & 0.2684  \\ \hline
LightGBM  & 99.885 & 99.885 & 0.6879  & 0.0226 & 126.7764  & 0.6879  \\ \hline
CatBoost  & 99.826 & 99.825 & 0.1809  & 5.9565 & 254.0400  & 0.1809  \\ \hline
LSTM      & 99.470 & 99.470 & 11.4647 & 0.3311 & 2937.1982 & 11.4647 \\ \hline
\end{tabular}}
\end{table}

Analysis of our original dataset revealed that the RF model achieved the highest accuracy and F1-score (99.895\%), followed closely by LightGBM (99.885\%) and XGBoost (99.835\%). From a computational perspective, XGBoost demonstrated balanced performance with moderate inference time (0.2684 seconds) and memory usage (274.62 KB), positioning it as a viable compromise between resource efficiency and predictive accuracy. CatBoost achieved the lowest accuracy (99.826\%) among traditional algorithms, while LSTM performed worst overall at 99.470\% accuracy with substantially higher resource demands (2937.19 KB memory, 11.4647 seconds inference time). Given our focus on resource-constrained environments, CatBoost and LSTM models were excluded from further analysis. LSTM’s underperformance, despite its architectural complexity, reinforced our decision to concentrate on tree-based ML approaches, which demonstrated a superior balance between efficiency and performance.

After applying data pre-processing steps—encoding, imputation, and normalization—we performed feature selection and data balancing. This reduced the dataset size by 87.3\% (from 22.06 MB to 2.8 MB), retaining key features (13 to 8 columns) and approximately one-third of instances (151,894 to 48,985 rows). The reduced dataset maintained model accuracy while significantly decreasing time and memory requirements across both training and test subsets. The optimized dataset enabled the implementation of pruning and quantization techniques for model complexity reduction. Table 2 compares the performance metrics of the three best-performing models before and after applying TinyML techniques.

Table 2 highlights the trade-off between model performance and computational efficiency achieved through the implementation of TinyML with XGBoost. While the standard XGBoost model slightly outperformed the TinyML-optimized version in terms of accuracy and F1-score (99.835\% versus 99.806\%), the TinyML approach delivered remarkable improvements in computational efficiency. Specifically, the inference time per sample was drastically reduced from 8.8357 milliseconds to 0.0045 milliseconds, and the cell execution time decreased significantly from 0.2684 seconds to 0.0368 seconds. Similarly, the memory usage showed a drastic reduction from 274.6221 KB to 7.4375 KB. These findings demonstrate that TinyML optimization offers substantial computational gains with only a negligible reduction in model performance (less than 0.03\% in accuracy). This makes it an ideal candidate solution for resource-constrained environments where efficiency is a critical consideration.

\begin{table}[t]
\centering
\caption{Comparison of ML models with and without TinyML}
\label{tab:tinyml-comparison}
\renewcommand{\arraystretch}{1.2}
\scalebox{0.63}{
\begin{tabular}{|>{\centering\arraybackslash}m{5em}|%
                >{\centering\arraybackslash}m{7em}|%
                >{\centering\arraybackslash}m{4.5em}|%
                >{\centering\arraybackslash}m{4.5em}|%
                >{\centering\arraybackslash}m{4.5em}|%
                >{\centering\arraybackslash}m{6.5em}|%
                >{\centering\arraybackslash}m{4.5em}|%
                >{\centering\arraybackslash}m{6em}|}
\hline
\multirow{2}{*}{\textbf{Model}\rule{0pt}{3.2ex}} &
\multirow{2}{*}{\makecell{\textbf{With or}\\\textbf{Without} \\\textbf{TinyML}}\rule{0pt}{3.2ex}} &
\multicolumn{2}{c|}{\textbf{Performance}\rule{0pt}{3.2ex}} &
\multicolumn{4}{c|}{\textbf{Computational Performance}\rule{0pt}{3.2ex}} \\ \cline{3-8}
& & \textbf{Accuracy (\%)} & \textbf{F1 score (\%)} &
\textbf{Inference time (sec)} & \textbf{Inference time (ms/sample)} &
\textbf{Memory usage (KB)} & \textbf{Cell execution time (sec)} \\ \hline
\multirow{2}{*}{RF}
  & Original    & 99.895 & 99.895 & 0.3181 & 0.0104 & 135.3506 & 0.3181 \\ \cline{2-8}
  & With TinyML & 99.765 & 99.765 & 0.0766 & 0.0078 & 149.2959 & 0.0766 \\ \hline
\multirow{2}{*}{XGBoost}
  & Original    & 99.835 & 99.835 & 0.2684 & 8.8357 & 274.6221 & 0.2684 \\ \cline{2-8}
  & With TinyML & 99.806 & 99.806 & 0.0633 & 0.0064 &   7.4375 & 0.0633 \\ \hline
\multirow{2}{*}{LightGBM}
  & Original    & 99.885 & 99.885 & 0.6879 & 0.0226 & 126.7764 & 0.6879 \\ \cline{2-8}
  & With TinyML & 99.694 & 99.693 & 0.1226 & 0.0125 &   7.8633 & 0.1226 \\ \hline
\end{tabular}}
\end{table}

\section{Conclusions}

This paper underscores the potential of TinyML-based solutions as a promising direction for securing critical RDS and IoT systems. By focusing on protecting RDSs from cyber threats, our research contributes to the broader goal of safeguarding critical infrastructure in diverse industries against emerging cybersecurity challenges. Our work has shown that the application of multiple TinyML techniques, including feature selection, hyperparameter tuning, sampling, pruning, and quantization, has successfully enhanced the computational efficiency of our XGBoost model while maintaining near-optimal accuracy (with less than 0.03\% reduction). These optimizations significantly reduced inference time and memory usage, making the model highly suitable for real-time applications in resource-constrained environments such as edge devices operating closer to RDSs. The results strongly support our initial assumption that a predictive ML model can serve as a viable and effective option for deploying an additional layer of security at the edge of critical architectures. 

\section{Acknowledgment}

This work is partially funded by the Natural Sciences and Engineering Research Council of Canada (NSERC). The work is also supported by the on-going International Atomic Energy Agency (IAEA) Coordinated Research Project (CRP) J02017 entitled “Enhancing Computer Security for Radiation Detection Systems.”

\end{document}